\documentclass[twocolumn,tighten,astrosymb,twocolappendix]{aastex7}

\usepackage{comment}
\usepackage{amsmath,amssymb,amstext}
\usepackage{subfigure}

\usepackage[version=4]{mhchem}
\usepackage{multirow}
\usepackage[utf8]{inputenc}
\usepackage{lmodern}
\usepackage{balance}
\usepackage{mathtools,nccmath}%

\usepackage{environ}

\usepackage{natbib}

\usepackage{xcolor}

\newcommand{\s}{\,s$^{-1}$}

\newcommand{\lx}{L_{\rm x}}

\newcommand{\rl}{\mathcal{R}_L}
\newcommand{\lb}{\ell_{\rm c}}
\newcommand{\lc}{\ell_{\rm c}}
\newcommand{\g}{\mathcal{G}}

\newcommand{\daniel}[1]{{\bf \color{teal} #1}}

\def\app#1#2{%
  \mathrel{%
    \setbox0=\hbox{$#1\sim$}%
    \setbox2=\hbox{%
      \rlap{\hbox{$#1\propto$}}%
      \lower1.1\ht0\box0%
    }%
    \raise0.25\ht2\box2%
  }%
}

\begin{document}

\title{\large On The Nonthermal Power Laws In Magnetized Turbulent Plasmas}

\author[orcid=0000-0001-9475-5292]{Rostom Mbarek}
\email[show]{rmbarek@princeton.edu}
\affiliation{Department of Astrophysical Sciences, Princeton University, Princeton, NJ 08544, USA}

\author[orcid=0000-0002-5408-3046]{Daniel Gro\v selj}
\affiliation{Centre for mathematical Plasma Astrophysics, Department of Mathematics, 
KU Leuven, B-3001 Leuven, Belgium}
\email{daniel.groselj@kuleuven.be}

\author[orcid=0000-0001-7801-0362]{Alexander Philippov}
\affiliation{Department of Physics, University of Maryland, College Park, MD 20742, USA}
\email{sashaph@umd.edu}

\begin{abstract}

Building on recent progress in the understanding of particle transport in magnetized plasmas, we derive a scaling law for the formation of nonthermal spectral tails in mildly and strongly magnetized turbulent environments. We validate this scaling using driven-turbulence particle-in-cell simulations that incorporate particle escape, allowing the system to reach a steady state. The simulation results show good agreement with our theoretical predictions. We then discuss the astrophysical implications of these findings, focusing on proton acceleration in the coronae of supermassive black holes and the resulting high-energy neutrino emission.

\end{abstract}

\section{Introduction}

Stochastic acceleration in turbulent astrophysical environments is a naturally occurring process that can explain the formation of nonthermal particle populations in various systems, such as kiloparsec-scale jets of active galactic nuclei \citep[e.g.,][]{kimura+18, mbarek+21a}, disk–jet systems in the vicinity of black holes  \citep[e.g.,][]{mahlmann+20,ripperda+20}, and pulsar wind nebulae \citep[e.g.,][]{begelman98,lyutikov+19,luo+20}. Understanding the mechanisms that govern particle acceleration in these environments is critical to unraveling the origins of high-energy phenomena. In such environments, turbulence spans a wide range of scales, where particles are stochastically energized through the Fermi mechanism \citep{Fermi49}.
This leads to energy distributions that significantly deviate from thermal equilibrium.  
A striking example is provided by strongly magnetized turbulent systems, where particles self-consistently develop extended power-law tails in their energy spectra, as shown by recent kinetic plasma simulations \citep[e.g.,][]{zhdankin+17,comisso+18,demidem+20,wong+20,pezzi+22,nattila+21}. \citet{lemoine+20} interpret these power laws as the result of a steady exchange between unaccelerated and accelerated particle populations, which maintains the nonthermal distribution.

Recent kinetic simulations highlight the role of
intermittent magnetic structures spanning a broad range of scales,
such as reconnecting current sheets \citep[e.g.,][]{mallet+17,boldyrev+17,loureiro+20, pezzi+22,xu+23,lemoine+24}, which enhance particle scattering.

\subsection{Outline}
In this Letter, we present an analytical framework for describing nonthermal particle distributions in magnetized turbulent media, circumventing some of the challenges of the traditional Fokker–Planck formalism. Our approach features a scale-dependent derivation of the particle acceleration time in both mildly and strongly magnetized turbulence, an analytical prediction of nonthermal spectral slopes, and validation via particle-in-cell (PIC) simulations of driven turbulent plasmas with large-amplitude magnetic fluctuations satisfying $\delta B / B \approx 1$. While our simulations are restricted to two dimensions, they span a large dynamic range and incorporate an escape 
mechanism that enables the system to reach a steady state without an energy pile-up.

We focus on plasmas with magnetization $\sigma = B^2 / (4\pi m n h c^2) \geq 0.1$, where $h$ is the specific enthalpy density, and $m$ and $n$ are the mass and number density of the plasma particles, respectively.  Our setup models particle distributions under the assumption of efficient acceleration facilitated by intermittent structures, which promote both scattering and energy gain. The resulting distributions identify the \emph{maximum power attainable at each energy scale}. These power levels are expected in the presence of intermittent structures with sharp magnetic field bends for Lorentz factors $\gamma > \sigma$, especially if radiative cooling is subdominant.

Finally, although our model is broadly applicable to various astrophysical sources, we specifically apply it to turbulent proton acceleration in the corona of the Seyfert galaxy NGC~1068—a promising candidate for $\gtrsim$ TeV neutrino emission \citep{IceCube-NGC1068}. Our predictions align well with current observational constraints in NGC 1068.

\subsection{Relevant Time Scales}

The analytical arguments in this Letter are based on a hierarchical ordering of relevant timescales. Particles first scatter off magnetic field bends in an energy-dependent manner, which produces a sustained average energy gain. As their Larmor radii grow, the corresponding acceleration time also increases and approaches the Alfvén crossing time. At this stage, interactions with magnetic field bends moving at the Alfvén speed ($v_A$) become less efficient, gradually shaping the overall particle distribution. Finally, a fraction of particles escapes the system, with the escape rate governed by the underlying scattering dynamics. The relevant timescales can be summarized as follows:

\paragraph{Scattering time ($t_{\rm sc}$)}
The scattering time is energy-dependent and governed by the intermittent structure of the turbulence \citep{lemoine23,kempski+23}. It sets the cadence of particle interactions with magnetic irregularities and plays a central role in both acceleration and transport.

\paragraph{Acceleration time ($t_{\rm acc}$)}
In a Fermi acceleration framework, particles gain or lose energy during each scattering event. Head-on collisions lead to energy gains, while tail-on collisions result in losses. Because head-on interactions are statistically favored, there is a net energy gain over time. Consequently, the acceleration time scales with the scattering time, yielding the relation $t_{\rm acc} \propto  t_{\rm sc}$.

\paragraph{Alfvén crossing time ($t_{\rm A}$)}
The Alfvén crossing time represents the timescale over which disturbances propagate across the system via Alfvén waves, making it the fundamental timescale for energy transport in magnetized plasmas. The interplay between $t_{\rm acc}$ and $t_{\rm A}$ at each scale sets the shape of the confined particle distribution.

\paragraph{Escape time ($t_{\rm esc}$)}
The escape time characterizes how long it takes for particles to diffusively exit the turbulent region. It depends on the scattering rate and particle energy. The energy spectrum of escaping particles differs from the confined ones, as lower-energy particles typically experience more difficulty escaping the system.

\section{Power Law Scaling in Magnetized Turbulence}\label{sec:analytics}

We derive a scaling law for the nonthermal energy distribution of stochastically accelerated particles 
in a magnetized source.
This scaling primarily depends on the magnetization $\sigma$,\footnote{For pair plasmas, $\sigma = \sigma_e$, where $\sigma_e$ is the electron magnetization. For electron-ion plasma, $\sigma = \frac{\sigma_e}{1 + \sigma_e/\sigma_p} \simeq  \sigma_p$, where $\sigma_p$ is the ion magnetization. For a pair plasma with an ion population, $\sigma \simeq \min(\sigma_e, \sigma_p)$, and ions are not energetically relevant if $\sigma_p > \sigma_e $, without accounting for the impact of cooling.} the coherence scale of the magnetic field $\lc$, and plasma skin depth $d_e$. 
The scaling can be readily applied to pure pair plasmas, electron-ion plasmas, and pair plasmas with a subdominant ion population ($\sigma_p < \sigma_e$) as well, since the only parameter that depends on the composition is $v_A$.

The scaling is relevant beyond an injection phase (defined below), characterized by $\gamma \simeq \sigma$, where $\gamma$ is the Lorentz factor of the particle. Assuming efficient acceleration, the scaling extends up to $\gamma \simeq \gamma_{\rm max}$, such that $\gamma_{\rm max} = eB \lc/mc^2$ corresponds to the maximum energy attainable \citep{hillas84} when the Larmor radius $\rl$ approaches the coherence length $\lc$ of the magnetic field, i.e., the scale of its largest bends.

\subsection{Bulk Power of the System}\label{sec:bulk}

During the initial energy boost, particle Lorentz factors remain limited to $\gamma \lesssim \sigma$, such that the Larmor radius
$\rl(\gamma) = {\gamma \beta m c^2}/{eB}$
satisfies $\rl \lesssim \sigma m c^2 / eB$, where $\beta$ is the particle velocity in units of $c$. This stage produces hard spectral slopes and extends up to $\gamma - 1 \simeq \sigma$ in magnetized systems \citep[e.g.,][]{sironi+14a,guo+14}. In highly magnetized environments ($\sigma \gg 1$), reconnection provides the dominant injection channel in decaying turbulence \citep{comisso+18,comisso+19}. Particles energized to $\gamma \sim \sigma$ then reach $\rl \sim d_e$, enabling interactions with intermittent magnetic structures on scales $l \gtrsim d_e$ (see \S~\ref{sec:interm}). In more moderate magnetizations, shocks could potentially mediate this initial boost \citep{groselj+26}.

Such a boost is fundamentally linked to the bulk power of the magnetized system $\mathcal{L}_b$, such that $\mathcal{L}_b \simeq \gamma_{\rm b} \bar{n} mc^2/\bar{t}_{\rm esc}$, where $\bar{t}_{\rm esc}$ is a typical escape time, $\bar{n}$ the bulk density (peak of the density distribution) of the system, and $\gamma_{\rm b}$ the Lorentz factor of the bulk density. The power $\mathcal{L}_b$ can be set by the energy balance between the escaping particle power and turbulence power per unit volume, $\mathcal{L}_b \approx \delta B^2 / (4\pi t_0)$, where $t_0$ is the turbulence cascade time. For $\delta B \approx B$, it is reasonable to assume that $\bar{t}_{\rm esc}\approx t_0 $ \citep[e.g., see][]{gorbunov+25}
and then $\gamma_{\rm b} \approx \sigma = B^2/(4\pi h\bar{n} mc^2)$. 
Consequently, the energy-dependent power of the system $ \mathcal{L}(\gamma) $ must satisfy $ \mathcal{L}(\gamma) \leq \mathcal{L}_b $ to ensure energy conservation, thereby constraining the particle distribution for $\gamma \geq \sigma$.

\subsection{Scale-Dependent Energy Conservation}\label{sec:interm}

For a particle with Larmor radius $\rl \sim l$, its Lorentz factor $\gamma_l$ can be associated with the spatial scale $l$, so that \emph{the energy and length scales relevant to gyration may be used interchangeably}. This is particularly useful because particles predominantly interact with structures of size $\rl \sim l$ \citep{lemoine23,kempski+23}. The same scale therefore also selects the fluctuations that dominate stochastic Fermi acceleration, since particles primarily exchange energy with structures satisfying $\rl \sim l$. In regions with $\sigma \gtrsim 0.1$, the fluctuation energy at scale $l$ is dynamically important, in the sense that electromagnetic stresses are no longer a small perturbation to the plasma dynamics. We therefore express the accessible energy at scale $l$ in terms of the magnetic fluctuation amplitude $\delta B_l$, which can be taken as an order-of-magnitude measure of the turbulent energy per scale. In relativistic, strongly magnetized turbulence, PIC simulations show that electric and magnetic fluctuations can become comparable,$\delta \mathcal{E}_l \sim \delta B_l$, and may dominate over bulk kinetic fluctuations \citep{vega+22}. Up to factors of order unity, $\delta B_l^2/8\pi$ is therefore taken to track the electromagnetic fluctuation energy available to resonant particles, with the replacement $\delta B_l^2 \rightarrow \delta B_l^2 + \delta \mathcal{E}_l^2$ appropriate in the strongly magnetized limit.

In the maximal-energization limit, particles with $\rl \sim l$ extract an order-unity fraction of the fluctuation energy at that scale, so the corresponding population of density $n_l$ and energy $E_l=mc^2\gamma_l$ satisfies the scale-dependent constraint
\begin{equation}\label{eq:enCons}
E_l^2\frac{dn_l}{dE_l} \simeq mc^2\gamma_l n_l \lesssim \frac{\delta B_l^2}{8\pi},
\end{equation}
with $\delta B_l^2/8\pi \rightarrow (\delta B_l^2+\delta \mathcal{E}_l^2)/8\pi$ when electric fluctuations contribute comparably. Equation~\eqref{eq:enCons} is thus a scale-dependent maximal-energization condition rather than a strict equipartition relation, providing an energetic upper bound on the particle energy density at a given scale, valid when particles remain confined long enough to sample an order-unity fraction of the turbulent electromagnetic energy.

For scales where $\delta B_l \sim B$,\footnote{The required volume fraction of regions with $\delta B \sim B$ is uncertain, but their influence should dominate over sufficiently long timescales. Additionally, scales with $\delta B_l \sim B$ likely extend to $l \ll \lc$ \citep{kempski+25}, highlighting the possibility of reaching the maximum power even at smaller scales.} the particle spectrum is then dominated by the highest-energy particles, yielding
\begin{equation}\label{eq:enCons-2}
E_l^2\frac{dn_l}{dE_l} \simeq mc^2\gamma_l n_l \simeq \frac{B^2}{8\pi} \simeq \frac{\bar n mc^2 \sigma }{2},
\end{equation}
where we invoke the bulk-energy argument of \S\ref{sec:bulk}, such that $\gamma_b \simeq \sigma$. The spectral slope for $\gamma_l \ge \sigma$ is thus constrained by the global energy budget, yielding ${dn_l}/{d\gamma_l} \propto \gamma_l^{-2}$.

However, this slope of $\propto \gamma_l^{-2}$ can only form if particles are efficiently accelerated at scale $l$ in regions where $\delta B_l \sim B$. As the Larmor radius $\rl$ (and therefore $\gamma_l$) increases and approaches the magnetic coherence length $\lc$, interactions with magnetic bends of size $l$ become less effective. 
Consequently, the maximum power attainable per scale decreases as $l \to \lc$, reflecting the reduced interaction probability.
We can rewrite Equation~\eqref{eq:enCons-2} as,
\begin{equation}\label{eq:enCons-3}
E_l^2\frac{dn_l}{dE_l} \simeq \frac{mc^2 \sigma }{2} \tilde{n}_l ,
\end{equation}
where $\tilde{n}_l \leq \bar{n}$ denotes the subset of particles that effectively interact with these accelerating regions as $l \to \lc$. A population satisfying ${dn_l}/{d\gamma_l} \propto \gamma_l^{-2}$ is recovered when $\tilde{n}_l=\bar{n}=\mathrm{constant}$. In this picture, Equation~\eqref{eq:enCons-2} applies for $l \ll \lc$ when structures with $\delta B/B \sim 1$ are present, while Equation~\eqref{eq:enCons-3} becomes appropriate in the limit $l \to \lc$.

\subsection{Slope Steepening at the Largest Scales}

Following \citet{bell78a} and \citet{lemoine21}, we can construct details of the power law governing stochastic acceleration in turbulent media.
For any statistical power-law-generating acceleration
mechanism, we can define a fractional energy gain per acceleration cycle $\mathcal{G}$, along with a probability $\mathcal{P}$ of remaining in the accelerating or confining region within one scattering time.
After $p$ acceleration cycles, there are $N(>E) = N_0 \mathcal{P}^p$ particles with energy $E = E_0 \mathcal{G}^p$. 
By eliminating $p$, and differentiating the cumulative number of particles $N(>E)$, we recover a power law distribution for the number of particles per unit energy, $ {d\tilde{n}_l}/{d\gamma_l} \propto \gamma_l^{-1+ \ln \mathcal{P}/\ln \mathcal{G}} $, resulting in $\tilde{n}_l \simeq \gamma_l {d\tilde{n}_l}/{d\gamma_l} \simeq \bar{n} \gamma_l^{\ln \mathcal{P}/\ln \mathcal{G}} $. Solving for the left-hand side of Equation~\eqref{eq:enCons-3}, which extends Equation~\eqref{eq:enCons-2} for $\gamma \to \gamma_{\rm max}$, yields the following\footnote{This is implied by both equations in regions where $\delta B / B \sim 1$.},
\begin{equation}\label{eq:generalEq}
    \frac{{\rm d}n_l}{{\rm d} \gamma_l} \simeq \frac{\sigma \bar{n}}{2} \gamma_l^{-2 + \ln \mathcal{P}/\ln \mathcal{G}}
\end{equation}
Therefore, if $\mathcal{P} \to 1$, then $\frac{dn}{d \gamma} \propto \gamma^{-2}$. Otherwise, the slope steepens according to $\mathcal{P}$ and $\mathcal{G}$. We can infer that scales characterized by magnetic structures with $\delta B / B \sim 1$ are smoothly connected, yielding a consistent power-law behavior as described by Equation~\eqref{eq:generalEq}. Equation~\eqref{eq:generalEq} should be interpreted as a differential contribution to the confined distribution at energy $\gamma_l$, associated with particles whose $\mathcal{R}_L\sim l$. Because the same scaling applies across scales, these local contributions form a single global power law.

\paragraph{The Probability $\mathcal{P}$ for Confined Particles}

The confinement probability $\mathcal{P}$, and thus the plasma population forming the power-law tail, evolves according to the relationship between the acceleration time $t_{\rm acc}$ and the Alfvén crossing time $t_A \approx \lc / v_A$. The Alfvén speed is defined as $v_A = c \sqrt{\sigma / (h + \sigma)}$. Since $t_{\rm acc} \propto t_{\rm sc}$, the acceleration time is both energy- and scale-dependent. As $\rl$ increases, so does $t_{\rm acc}$. When $t_{\rm acc} \to t_A$, where $t_A$ can be thought of as the timescale over which scale- and energy-dependent bends propagate along magnetic field lines, $\mathcal{P}$ decreases, leading to a steepening of the distribution $\frac{dn}{d\gamma}$.

The timescale $t_A$ is expressed in terms of the coherence length $\lc$ because the largest possible size of the accelerating region is associated with the largest magnetic field bends in the system. The evolution of the nonthermal particle population can be modeled as a Poisson process, ${dn_c}/{dt_{\rm acc}} = -n_c/t_{\rm A}$, where $1/t_A$ is the rate of interactions for the acceleration timescale $t_{\rm acc}$, and $n_c$ denotes the number density of confined particles.

The solution to this process can be written as $ n_c  \propto e^{-t_{\rm acc} / t_A} $, and the probability of remaining confined within the accelerating region becomes $ \mathcal{P} = e^{-t_{\rm acc}/t_{\rm A}} $, with $t_{\rm acc} \in [0, \infty)$. The effect of $\mathcal{P}$ becomes significant primarily at the highest energies ($\gamma \to \gamma_{\rm max}$), while for lower energies, $\ln \mathcal{P} \approx 0$.

\paragraph{The Energy Gain $\mathcal{G}$} 

For magnetized turbulent systems, \citet{vega+24} emphasized the importance of the relative strength between turbulent fluctuations $\delta B$ and the guide field $B$. In regimes where $\delta B/B \ll 1$, curvature drift plays a major role, and consequently, curvature acceleration dominates the particle energization process. This mechanism remains efficient even for small pitch angles. In this limit, $E_l^2\frac{dn_l}{dE_l} \simeq mc^2 \gamma_l n_l \simeq \delta B_l^2/8 \pi \ll B^2/8\pi$, and therefore, the assumptions from \S\ref{sec:interm} do not hold. The resulting particle spectrum in this case can be quite hard, with a slope consistent with $\frac{{\rm d}n_l}{{\rm d} \gamma_l} \propto \gamma_l^{-1 + \ln \mathcal{P}_{\rm cu}/\ln \mathcal{G}}$, where $\mathcal{G} = 1+ \Delta \gamma/\gamma \simeq \pi (v_A/c) (1 + (B/\delta B)^2)^{-1/2}$ \citep{vega+24}, and $\mathcal{P}_{\rm cu}$ could depend on the local plasma properties \citep[e.g.,][]{lemoine21}. In general, the net energy gain remains small, particularly when $\delta B/B \leq 0.1$.

For more moderate guide fields with $\delta B/B \sim 1$, mirror acceleration can dominate over curvature acceleration as a result of enhanced pitch-angle scattering \citep{vega+24}. Mirror acceleration depends on the energy-dependent average pitch angle $\theta$ of the particle distribution, such that $\Delta \gamma/\gamma \simeq 2 (v_A/c) \cos{\theta}$, where $v_A$ is the mildly relativistic Alfvén velocity of the plasma. The corresponding energy gain can then be estimated as $\mathcal{G} \simeq 1 + 2 (v_A/c) \langle \cos{\theta} \rangle$, where $\theta$ follows a nearly uniform distribution for $\delta B/B \sim 1$ \citep[Figure~6 in][]{vega+24}. In this case, the average pitch angle factor is $\langle \cos{\theta} \rangle = \frac{1}{\pi} \int_{-\pi/2}^{\pi/2} \cos\theta  d\theta \simeq 2/\pi$ for $\theta \in [-\pi/2, \pi/2]$.

A self-consistent analysis of the dominant acceleration mechanisms in the simulations, together with the corresponding pitch-angle distributions, is beyond the scope of this study but would be valuable. Overall, curvature acceleration is expected to dominate at scales $l \ll \lc$, where it can efficiently pre-energize particles and inject them into regions where $\delta B \sim B$. At such scales, particles undergo large pitch-angle scattering and can experience both mirror and curvature acceleration, enabling them to reach the maximum attainable power at each scale. However, intermittent structures could also play a role in accelerating particles through mirror acceleration at $l \ll \lc$.
For the remainder of this paper, we focus on mirror acceleration, as it is dominant for $\delta B \sim B$.

\paragraph{The Acceleration Time $t_{\rm acc}$} 
For Fermi processes, we can compute the acceleration time of particles $t_{\rm acc} =\gamma^2/D_{\gamma \gamma}$, where $D_{\gamma \gamma} \simeq (\Delta \gamma)^2/(2 \Delta t)$ is the energy diffusion coefficient \citep[e.g.][]{lemoine19,lemoine25}. Based on our energy gain discussion, we express $(\Delta \gamma)^2 \simeq (4\gamma^2/\pi)(v_A/c)^2 $, resulting in $D_{\gamma \gamma} \propto \sigma$ for $\sigma \sim 1$ \citep[consistent with][]{wong+25}. For Fermi acceleration, the canonical time $\Delta t$ is the scattering time $t_{\rm sc}$ since acceleration occurs when particles are scattered. The scattering time $t_{\rm sc}$ can be expressed as $t_{\rm sc} \simeq \lb^{1-r} \rl^{r}/c$ \citep{lemoine23,kempski+23}.
The exponent $r$ is obtained from $\mathcal{P}_\kappa$, the probability distribution of $\langle l \kappa_l \rangle$, where $\langle \kappa_l \rangle$ is the averaged magnetic field curvature, and $l$ is a scale below $\lc$. 
The distribution $\mathcal{P}_\kappa$ exhibits power-law tails, especially in the presence of intermittent structures at small scales, $\mathcal{P}_\kappa \propto ( l \kappa_l )^{-\alpha}$. The slope $\alpha$ of such tails sets the properties of the system for $l \kappa_l  \gtrsim 1$, and the value of $r$ depending on the cascade \citep{lemoine23}. 

We finally obtain for the acceleration time,
\begin{equation}\label{eq:tacc}
	t_{\rm acc} \simeq \frac{2\gamma^2}{(\Delta \gamma)^2} \frac{\lc^{1-r} \rl^{r}}{c}  \simeq \frac{\pi}{2c} \left( \frac{h + \sigma_p}{\sigma_p} \right) \lb^{1-r} \rl^{r} 
\end{equation}
suggesting that the acceleration time can be energy-dependent if $r > 0$. Note that the right-hand side of Equation~\eqref{eq:tacc} is relevant for flat pitch angle distributions in $\delta B\sim B$ regions.

\paragraph{Spectral slope of the confined distribution}
We eventually obtain an expression for the confined plasma following Equation~\eqref{eq:generalEq}, such that $ \frac{dn_c }{d\gamma} = \frac{\sigma \bar{n}}{2} \gamma^{-s}$. The spectral slope is expressed as $s = 2 - \ln \mathcal{P}/\ln \mathcal{G}$, yielding,
\begin{equation}\label{eq:slope}
	s \simeq 2 + \frac{\pi}{2} \sqrt{\frac{h + \sigma_p}{\sigma_p}} \left(\frac{\gamma \frac{m_p}{m_e} d_e}{\sqrt{\sigma_e} \lc} \right)^{r} \frac{1}{\ln \mathcal{G}}
\end{equation}
where $\rl  = \gamma \frac{m_p}{m_e} d_e/\sqrt{\sigma_e}$, for a species of mass $m_p$, and magnetization $\sigma_p = \sigma_e m_e/m_p$. Considering the large pitch angle scattering expected for $\rl \sim l$, we adopt the mirror acceleration estimate discussed above, $\mathcal{G} \simeq 1 + (4/\pi)(v_A/c)$, and we basically recover an expression for the slope that depends on macro plasma properties of the system, i.e., the magnetization and coherence scale of the system in units of skin depth. 

In the remainder of this paper, we test properties of the confined distribution with PIC simulations and compare with the above scaling. Overall, this calculation implies that for magnetized turbulent systems with magnetization $\sigma \geq 0.1$, particles are likely accelerated with a slope $s \to 2 $ if $\delta B/B\sim1$ regions are sustained at scales $\rl \sim l \ll \lc$. If intermittency is not sustained at such intermediate scales, other scenarios are needed \citep[e.g.,][]{mbarek+24,lemoine+24}.

\paragraph{Diffusive Escape}
For accelerated particles in the power-law tail that remain within the system of size $S$ within a timescale $t_A$, the probability of retention follows the equation $\frac{dn}{dt_A} = -n/t_{\rm esc}$. Solving this equation gives $ n \propto e^{-t_A/t_{\rm esc}} $. 
We can then express the probability of escape as $\mathcal{P_{\rm esc}} = 1 - e^{ - t_A/t_{\rm esc}}$.
Moreover, the escape time $t_{\rm esc}$ is set by a random walk, $ t_{\rm esc} \simeq S^2/(c^2 t_{\rm sc}) $. 
We obtain $t_{\rm esc} \simeq S^2 / (\lc c)(\rl/\lc)^{-r} = S^2 / (\lc c)(\gamma/\gamma_{\max})^{-r}$, consistent with measurements in driven 3D turbulent pair plasma PIC simulations with diffusive escape \citep{gorbunov+25}, where an escape time of $t_{\rm esc} \propto (\gamma/\gamma_{\max})^{-0.3}$ is extracted for $r \simeq 0.3$.

The escaping population can finally be expressed as,
\begin{equation}\label{eq:escape}
\begin{split}
    \frac{dn_{\rm esc}}{d\gamma} = \frac{\sigma \bar{n}}{2} \gamma^{-s}(1 - e^{ - t_A/t_{\rm esc}}) \\ \simeq \frac{\sigma \bar{n}}{2} \frac{\lc^2}{S^2} \sqrt{\frac{h+\sigma_p}{\sigma_p}} \left( \frac{\frac{m_p}{m_e} d_e}{\sqrt{\sigma_e} \lc} \right)^{r} \gamma^{-s+r}
\end{split}
\end{equation}
where $(1 - e^{ - t_A/t_{\rm esc}}) \simeq t_A/t_{\rm esc}$ for $t_A/t_{\rm esc} \ll 1$. We note that the slope is consistently harder than the confined one by a factor of $r$ (in agreement with \citealt{gorbunov+25} for $r\simeq 0.3$).

When considering the impact of escape, it is useful to separate regimes by timescales. When $t_{\rm esc} \gg t_A$ and $t_{\rm esc} \gg t_{\rm acc}$, escape does not control the in-situ power law, so the intrinsic spectrum remains $n_c \propto \gamma^{-s}$. If instead $t_A \gtrsim t_{\rm esc}$, then $1 - e^{-t_A/t_{\rm esc}} \to 1$, the escaped sample from the accelerator inherits the same slope $s$, while the in-situ spectrum is softened by escape at the rate $\propto \gamma^{r}$, giving $n_c  \propto \gamma^{-(s+r)}$ in steady balance.

\section{Comparison with kinetic simulations}
Considering the general nature of the results associated with Equation~\eqref{eq:slope}, we can test its general validity with a comparison with the power laws of nonthermal tails of confined populations in turbulent kinetic PIC simulations \footnote{A more detailed treatment could follow individual particles and compute their interaction probabilities explicitly, but this is beyond the scope of the present study.}. 
We choose a driven electron-ion turbulent plasma setup with magnetization $\sigma = \sigma_e/(1 +{\sigma_e}/{\sigma_{p}}) \simeq  \rm min(\sigma_p, \sigma_e) \sim 1$ with a guide field satisfying $\delta B/ B \simeq 1$, likely relevant for coronae and jets \citep{mbarek+24}. Throughout, the temperature is $k_b T/m_e c^2 = 0.002$, thus setting $h \simeq 1$.

We first extract the exponent $r$ directly from the simulations, then based on the pre-set values of $\sigma_p$, $\sigma_e$, and $\lc$, we retrieve the expected nonthermal slopes based on Equation~\eqref{eq:slope} and compare them with PIC results.

\subsection{Simulation Setup}

We perform a set of kinetic turbulence simulations using the PIC code \textsc{Tristan-MP v2} \citep{tristanv2}.
We set up a 2D computational domain of size $L^2$ and fill it with electron-ion plasma, with mass ratio of $m_p/m_e = 5$. The plasma skin depth $d_e$ is resolved with 1.5 cells and our time step $\Delta t = 0.25 \Delta x/c$. We drive a turbulent state on the box scale by imposing a time varying external current \citep{tenbarge+14}, 
thereby exciting Alfv\'enic perturbations.
The frequency and decorrelation rate associated with the driving are set as $\omega_0 = 1.4 (2 \pi v_A / L)$, and $\gamma_0 = 0.5 \omega_0$, respectively, with $v_A = c \sqrt{\sigma_p/(1 + \sigma_p)}$.

We consider periodic boundary conditions in all directions but allow particles to escape the simulation box once their displacement exceeds $L/4$ \citep[e.g.,][]{gorbunov+25}, in order to prevent energy pile-up in the system \citep[e.g.,][]{zhdankin21}. Specifically, we track each particle’s displacement, and when it reaches $L/4$, the particle is removed and replaced by a new one drawn from a thermal Maxwellian distribution. This escape criterion is motivated by the fact that the largest-scale magnetic structures in our simulations, characterized by the coherence length, are approximately $\lc \simeq L/4$. The escape condition is evaluated every 10 time steps. To mimic the conditions of strongly turbulent astrophysical systems, we also impose a mean guide field in the $z$-direction such that $\delta B / B = 1$.
We emphasize that particle escape in our simulations is an \emph{imposed} boundary condition: particles are removed once their displacement exceeds $L/4 \simeq \ell_c$, and replaced by fresh thermal particles. This condition is motivated by the theoretical picture of diffusive escape from a region of size $\ell_c$, as discussed in \S\ref{sec:analytics}, but is imposed explicitly rather than arising self-consistently from the dynamics.

We note that the electron inertial scale is resolved with 1.5 cells. This could limit the fidelity with which electron-scale injection physics can be captured. However, the nonthermal tail discussed here is shaped primarily during the subsequent stochastic acceleration stage, when particles have already been injected to $\gamma \sim \sigma$ and interact with structures on scales $\mathcal{R}_L \gtrsim d_e$ that are well resolved across the simulation dynamic range. We therefore expect the reported spectral slopes to be robust. We performed additional simulations with 4 cells per $d_e$ and obtained results similar to our large-size simulations with 1.5 cells per $d_e$.

\begin{figure}
	\centering
\includegraphics[width=0.48\textwidth,clip=false,trim= 0 0 0 0]{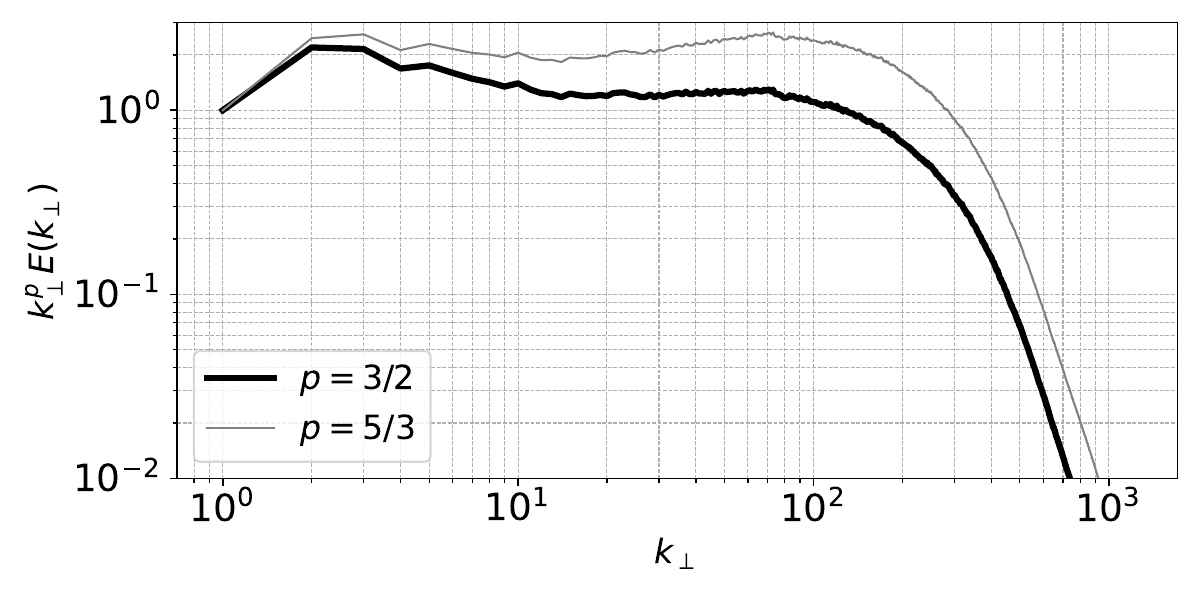}
	\caption{Power spectrum $E(k_{\perp})$ of the magnetic field for a simulation with $L/d_e = 3200$ and $\sigma_p = 2$. $E(k_{\perp})$ is scaled for reference by $k_{\perp}^p$, for $p = 3/2$ and $p = 5/3$. The spectrum is averaged for $tc/L \in [8, 10]$ after steady state is achieved.
    }
	\label{fig:pwr-spec}
\end{figure}

\subsection{Properties of the turbulent box}\label{sec:prop-turb-box}

\paragraph{\textbf{Magnetic power spectrum}}
We compute the magnetic power spectrum $E(k_{\perp})$ from the discrete Fourier transform of the fluctuating magnetic field in our simulations, such that, $E(k_{\perp}) dk = \sum_{k \in dk} \frac{\mathbf{B_k}.\mathbf{B^*_k}}{8 \pi}$. The magnetic power spectrum exhibits a scaling of, $E(k_{\perp}) \propto k_{\perp}^{-3/2}$. This scaling is consistent with results from high-resolution incompressible MHD simulations \citep[e.g.,][]{maron+01, cho+02}, as well as from large-scale 3D studies of compressible magnetized turbulence \citep{beattie+25}, and relativistic resistive MHD simulations with guide fields satisfying $\delta B/B  = 1/3$ \citep{chernoglazov+21}. This is generally consistent with \citet{boldyrev05,boldyrev06}'s dynamic alignment argument for strong magnetization and guide fields, where turbulent eddies stretch along the guide field, becoming increasingly elongated at smaller scales.

\paragraph{\textbf{Magnetic field curvature}}
The field-line curvature is $\boldsymbol{\kappa}=(\boldsymbol{b}\cdot\nabla)\boldsymbol{b}$, where $\boldsymbol{b}$ is the unit magnetic-field vector; its magnitude $\kappa$ quantifies the turning per unit arc length. To isolate contributions near a perpendicular scale $l\sim 1/k_\perp$, let $\tilde{B}_l$ denote the coarse-grained field at scale $l$ and $\delta B_l$ the associated fluctuation. The bend angle satisfies $\Delta\theta_l\sim \delta B_l/\tilde B_l$ for small angles, and the bend extends a distance $l_{\parallel}(l)$ along the local mean field, so the curvature contributed by that eddy is $\sim \Delta\theta_l/l_{\parallel}(l)$. With $\delta B_l/\tilde B_l\simeq 1$, this gives $\langle\kappa_l\rangle\sim 1/l_{\parallel}(l)$ (see Appendix~\ref{app:deltaB}), which we take as the average local curvature at scale $l$ over the limited dynamic range considered. For a dynamically aligned cascade consistent with $E(k_{\perp})\propto k_{\perp}^{-3/2}$, critical balance with alignment yields $l_{\parallel}(l)\sim \ell_c (l/\ell_c)^{1/2}$ \citep{boldyrev06}, hence $\langle \kappa_l \rangle \sim \ell_c^{-1}(l/\ell_c)^{-1/2}$. We obtain,
\begin{equation}\label{eq:kappa_l}
\langle \kappa_l \rangle \ell_c \sim \left(k_{\perp}\ell_c\right)^{1/2} \sim \left(l/\ell_c\right)^{-1/2}.
\end{equation}

\paragraph{\textbf{Statistics of magnetic field curvature}}
Following the methods presented in \citet{lemoine23}, we extract the distribution of magnetic field curvature strength $\kappa_l$ on scales $l \leq \lc$ in our PIC simulations. 
For a given scale $l$, we extract the coarse-grained\footnote{Coarse-graining attenuates high-frequency components while preserving the overall magnetic structures. It is applied using Gaussian smoothing with a standard deviation of extent $l$.} magnetic field $\Tilde{B}_l$ on scale $l$, and compute the curvature,
\begin{equation}\label{eq:kappa}
     \kappa_l  \simeq \frac{\Tilde{B}_l \times (\Tilde{B}_l \cdot \nabla) \Tilde{B}_l}{|\Tilde{B}_l|^3} \Bigg( \frac{\langle \Tilde{B} \rangle^{1/2} }{\Tilde{B}_l} \Bigg)
\end{equation}
such that $\langle \Tilde{B} \rangle^{1/2}$ is also averaged on $l$. 

\begin{figure}
	\centering
\includegraphics[width=0.48\textwidth]{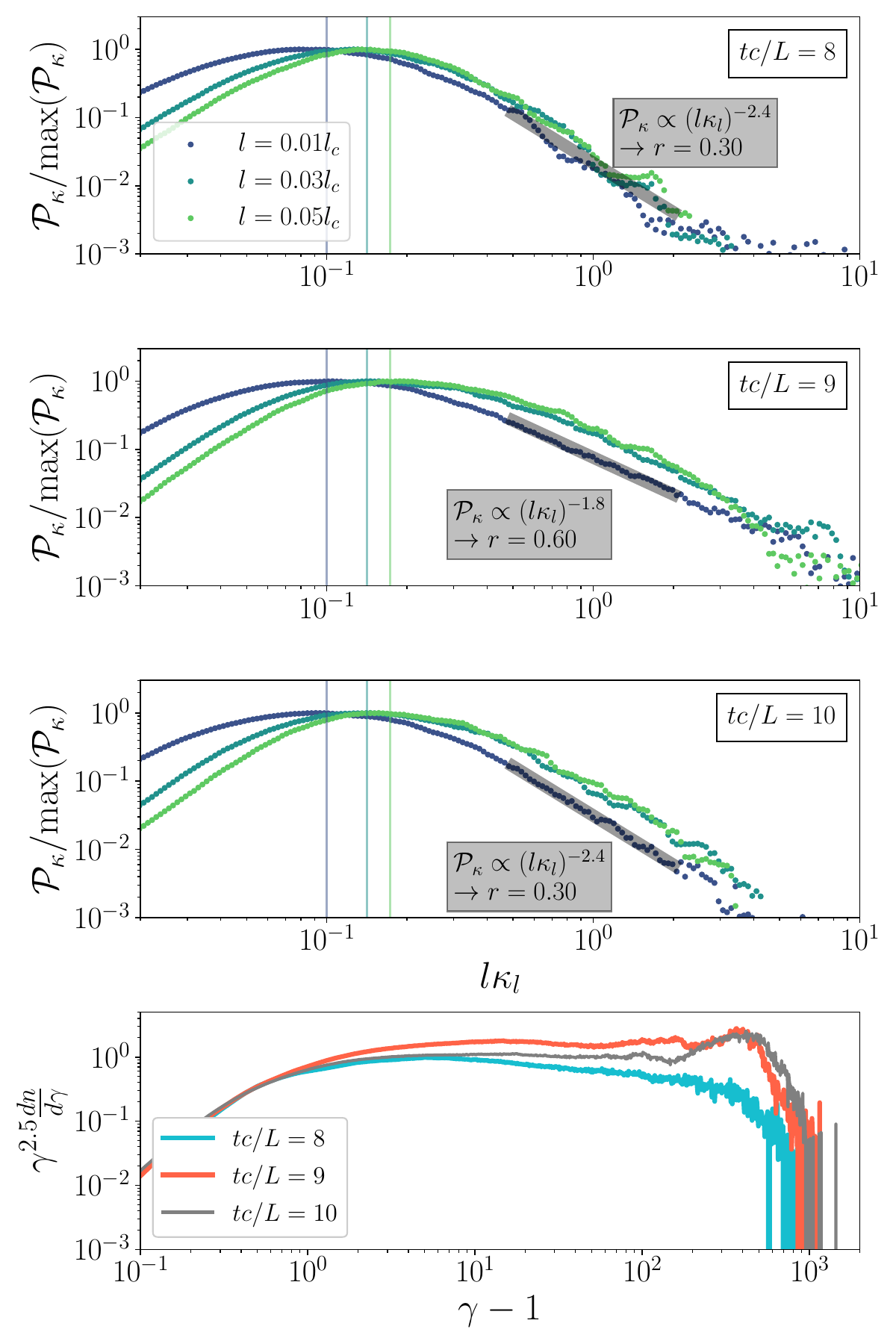}
	\caption{Upper Panels: Evolution of the curvature scale statistics in a 2D box of size $L/d_e = 3200$ with $\sigma_p = 2$, shown at different times. The curvature $\kappa_l$ is defined in Equation~\eqref{eq:kappa}. The resulting distributions exhibit power-law tails, scaling as $\mathcal{P}_\kappa \propto (l \kappa_l)^{-\alpha}$ for $l \kappa_l \gtrsim 1$. This power-law behavior reflects the underlying stochasticity of the system, which plays a direct role in shaping the particle spectra. Effectively, larger values of $r$ correspond to harder spectral slopes. The bottom panel illustrates the immediate influence of $\mathcal{P}_\kappa$ on the resulting particle spectra. The power-law slopes $\mathcal{P}_\kappa \propto (l\kappa_l)^{-\alpha}$ shown are best fits over the range $l\kappa_l \in [0.5, 2]$ and should be interpreted as indicative reference values. The slope varies somewhat with the fitting range and across realizations, reflecting the stochastic nature of the intermittent structures.
    }
	\label{fig:intermittency}
\end{figure}

\paragraph{\textbf{Extracting the coefficient $r$ in Equation~\eqref{eq:slope}}}
After the plasma reaches a steady state\footnote{Steady state is defined such that $\dot{U} \approx 0$, where $U$ is the total energy density of plasma. It is achieved for $t c/L \gtrsim 6$.}, we show the probability distributions $\mathcal{P}_\kappa$ of $ l \kappa_l $ for $\delta B/B =1$, $\sigma_p = 2$, and $L/d_e = 3200$ in the upper three panels of Figure~\ref{fig:intermittency}, as an example. The $\mathcal{P}_\kappa$ distributions exhibit i) peaks at $\sim (l/\lc)^{1/2}$, consistent with Equation~\eqref{eq:kappa_l}, and ii) a power law beyond the peak with a slope $\alpha \approx 2.4$ for $ l \kappa_l \gtrsim 1$. We note that the power-law tails are inferred over the limited range $l\kappa_l \in [0.5, 2]$, and the inferred $\alpha$ carries some uncertainty as a result.

For $\rl \sim l$, the mean free path for a scattering event $c t_{\rm sc}$ can be expressed based on the filling factor of regions satisfying $l \kappa_l \gtrsim 1$, such that $c t_{\rm sc} \sim l/\int_1^{\infty}\mathcal{P}_\kappa dx$, where $x = l\kappa_l$ \citep{lemoine23}. The $\mathcal{P}_\kappa$ distribution can in turn be expressed as $\mathcal{P}_\kappa \sim \langle l \kappa_l \rangle^{-1} (x/\langle l \kappa_l \rangle)^{-\alpha}$, where $\langle l \kappa_l \rangle$ corresponds to the peak of the distribution. Evaluating the integral for $\alpha >1$ and using Equation~\eqref{eq:kappa_l},
\begin{equation}\label{eq:scat_alpha}
    c t_{\rm sc} \simeq (\alpha - 1) l \langle l \kappa_l \rangle^{1 - \alpha} \sim \rl (\rl/\lc)^\frac{1 - \alpha}{2} = \rl^\frac{3 - \alpha}{2} \lc^\frac{\alpha - 1}{2}
\end{equation}
The exponent $r$ then becomes $r = \frac{3 - \alpha}{2}$, resulting in an average value of $r \approx 0.3$. However, harder $\mathcal{P}_\kappa$ spectra can be sustained for short periods of time, directly affecting the value of $r$, and thus the particle distribution. We note that Equation~\ref{eq:scat_alpha} differs from \citet{lemoine23}'s results because $E(k_{\perp}) \propto k_{\perp}^{-3/2}$.

In the bottom panel of Figure~\ref{fig:intermittency}, we show the particle spectra associated with each time $tc/L$.
We note i) the presence of quite hard power laws, ii) that changes in $r$ result in concurrent changes in the particle distribution, and iii) particles are energized up to the Hillas limit in our setup $\gamma_{\rm max} \gtrsim 10^3$.
The stochastic nature of the power-law in $\mathcal{P}_\kappa$-distributions influences particle spectra, with harder power laws in $\mathcal{P}_\kappa$ leading to correspondingly harder particle spectra. This variability likely originates from the intermittent formation of magnetic structures, which modulate scattering and acceleration probabilities. Such intermittency may be more pronounced in 2D simulations than in 3D setups.
Overall, the power law is shaped by the interaction with intense structures as seen in \citet{lemoine22}.

A more exact value for the slope $\alpha$ may become clearer with very large 3D simulation domains, which we leave for future study. Finally, we note that this type of stochasticity could result in variability in emission from magnetized turbulent astrophysical environments.
In the following, we further compare our Equation~\eqref{eq:slope} scaling with slopes from kinetic simulations.

\begin{figure}
	\centering
\includegraphics[width=0.48\textwidth,clip=True,trim= 0 0 0 0]{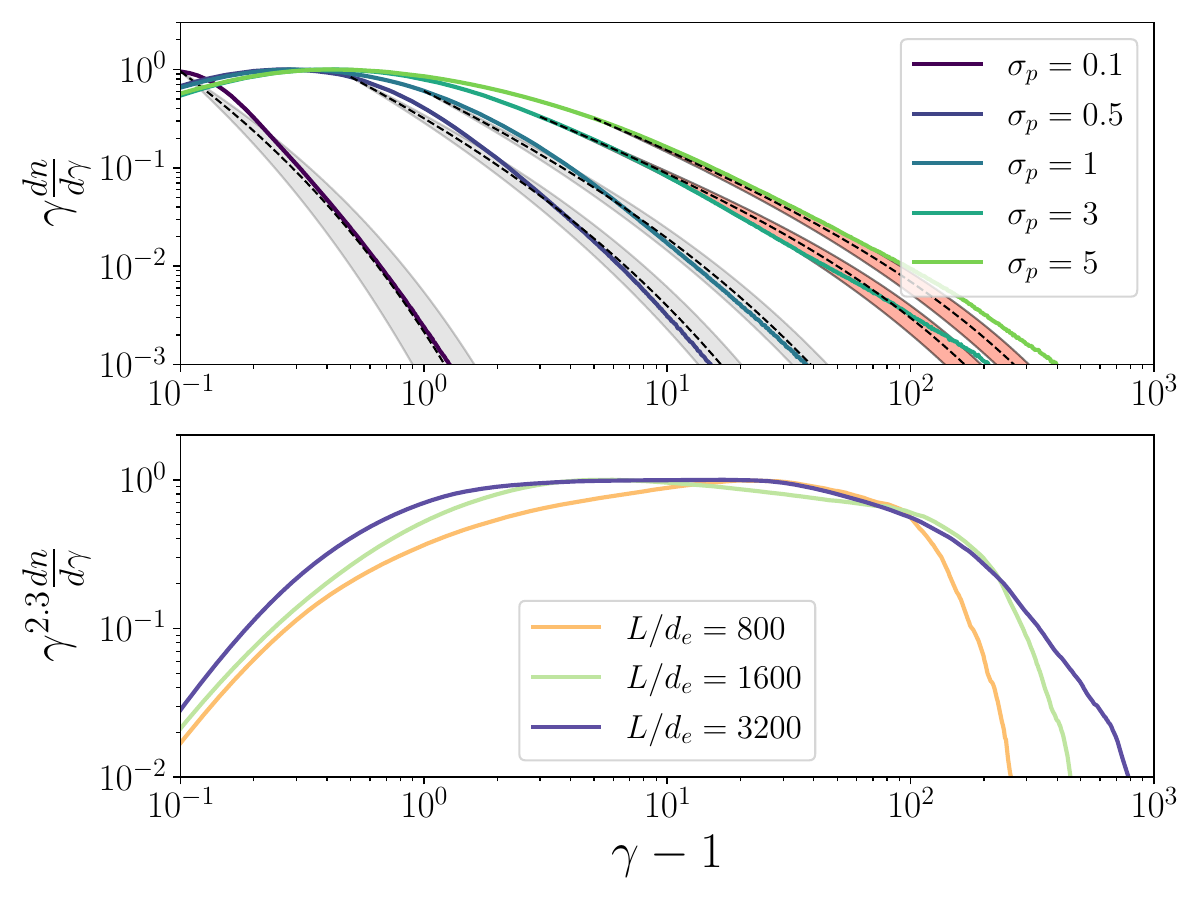}
	\caption{Upper panel: nonthermal slopes for different $\sigma_p$ initializations, compared with the analytical prediction of Equation~\eqref{eq:slope} with different r-values (dotted and gray region). The grayed regions are for $r =0.3\pm0.05$ and the reddened regions are for $r =0.5\pm0.1$.  
    Lower panel: nonthermal slopes for increasing box size (or increasing $\lc$) for $\sigma_p = 2$.
    While the slope tends towards $s = 2.3$ for the largest $L/d_e$, we expect $s \to 2$ for larger boxes, and thus astrophysical systems. All spectra in both panels are averaged for $tc/L \in [8, 10]$ after steady state is achieved.
    }
	\label{fig:comparison}
\end{figure}

\subsection{Comparison with slopes in kinetic simulations for $\rl \lesssim \lc$}\label{sec:compare}

In the upper panel of Figure~\ref{fig:comparison}, we present proton spectra from a set of simulations with box size $L/d_e = 1600$, $\lc \simeq L/4$, and mass ratio $m_p/m_e = 5$, for varying proton magnetizations $\sigma_p$. The spectra are averaged over the interval $ct/L = 8$ to $ct/L = 10$. To enable meaningful comparisons with the theoretical model, the PIC spectra are averaged over at least one characteristic Alfvén timescale, $t_A$, to allow the particle distribution to reach equilibrium. All spectra correspond to a phase in which the plasma has reached a steady state, characterized by $\dot{U} \approx 0$.
We find that the nonthermal tail emerges at $\gamma \gtrsim \sigma_p$, and for $\gamma \gtrsim \sigma_p$, the spectral slope becomes progressively harder with increasing magnetization.

In the upper panel of Fig~\ref{fig:comparison}, we fit our PIC results with Equation~\eqref{eq:slope} based on $r$-values extracted from Figure~\ref{fig:intermittency}. We find good agreement between the scaling in Equation~\eqref{eq:slope} and the slopes from our simulations for $r  = 0.3 \pm0.05$ if $\sigma_p \leq 1$, and for $r  = 0.5 \pm0.1$ if $\sigma_p \geq 3$.
For lower values of $\sigma_p$, we note a \textit{thermal bump} at lower energies $\gamma \sim \sigma_p$ signaling the impact of thermal particles on the slope. For larger values of $\sigma_p$, larger values of $r$ are necessary to fit the slopes because harder $\mathcal{P}_\kappa$ spectra are sustained for a longer time (see Appendix~\ref{app:intermit} for examples). A more extensive study on the impact of simulation initializations, including magnetization and 3D-effects, on $\mathcal{P}_\kappa$ is necessary, but is beyond the scope of the current Letter.

We note that the values of $r$ used in the upper panel of Figure~\ref{fig:comparison} are obtained by time-averaging $\alpha$ over the interval $ct/L \in [8, 10]$ and applying Equation~\eqref{eq:scat_alpha}. Because Equation~\eqref{eq:slope} is nonlinear in $r$, the time-averaged $r$ is not strictly equivalent to an average over instantaneous slopes. However, the variability is modest for $\sigma_p \leq 1$ (yielding $r \approx 0.3 \pm 0.05$) and somewhat larger for $\sigma_p \geq 3$ (yielding $r \approx 0.5 \pm 0.1$), where harder $\mathcal{P}_\kappa$ tails are sustained for a longer fraction of the averaging window (see Appendix~\ref{app:intermit}).

In the lower panel of Figure~\ref{fig:comparison}, we examine the impact of box size, used here as a proxy for the acceleration region scale $\lc$, on the particle spectra. As anticipated from Equation~\eqref{eq:slope}, larger boxes yield harder spectra at high energies, with spectral indices approaching $s \to 2$ as $\lc/d_e$ reaches more realistic values. Notably, the largest simulation box sustains a hard spectrum with $s \simeq 2.3$ even for particles with Larmor radii $\rl \sim l \ll \lc$, suggesting that a slope near $s \simeq 2$ could be expected at these scales.
At lower Lorentz factors, $\gamma \gtrsim \sigma$, the particle distribution is increasingly influenced by the thermal component, particularly in smaller boxes, which can artificially flatten the spectrum and make it appear harder than expected.

Although our simulations are constrained by a limited dynamic range and do not fully capture the $\rl \ll \lc$ regime, Figure~\ref{fig:comparison} indicates that in this limit, $s \simeq 2$, as the probability of particle confinement increases despite efficient acceleration. This aligns with the arguments in \S\ref{sec:analytics}, where we show that the total system power imposes an upper limit on the energy density of accelerated particles, ultimately shaping the overall particle distribution.

\begin{figure}
	\centering
\includegraphics[width=0.48\textwidth,clip=false,trim= 0 0 0 0]{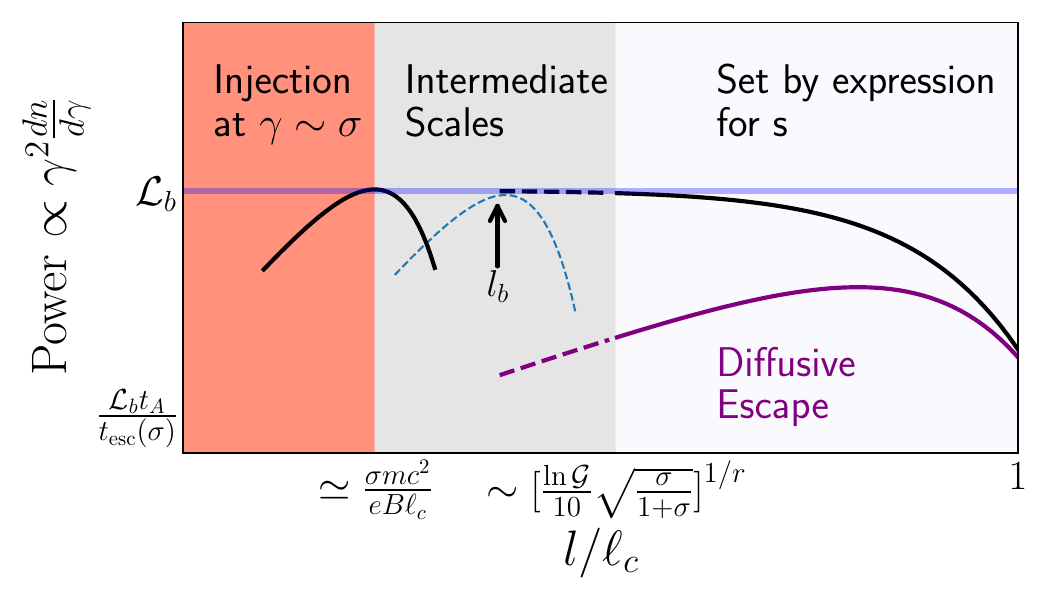}
	\caption{Depiction of the expected confined and escaping particle distributions at different scales satisfying $\rl \sim l$. For the confined distribution, injection occurs with a hard spectrum $s_{\rm inj}  = 1$. At scales $l \geq l_b$, regions where $\delta B/B \sim 1$ emerge, allowing particles to be accelerated and driving the spectrum toward a slope of $s = 2$ over sufficiently long timescales, even when such regions occupy only a small fraction of the volume (see \S\ref{sec:interm}). For efficient acceleration, Equation~\eqref{eq:slope} is relevant on scales $l > \frac{\sigma m c^2}{e B \lc}$. As for the escaping distribution, it is set by Equation~\eqref{eq:escape}. Generally speaking, our PIC simulations mostly probe scales beyond $\gtrsim \left(\frac{\ln \mathcal{G} }{10} \sqrt{\frac{\sigma}{1+\sigma}}\right)^{1/r} $ because of the limited dynamic range.
    }
	\label{fig:scales}
\end{figure}

\subsection{Synthesis}
Based on the analytical scalings presented in \S\ref{sec:analytics} and their comparison with PIC results, Figure~\ref{fig:scales} illustrates the expected particle distributions in magnetized turbulent systems with $\sigma \geq 0.1$. The scale $l_b$ marks the onset where regions with $\delta B/B \sim 1$ attain a non-negligible filling factor, enabling efficient particle acceleration. Particles are (i) injected through processes such as magnetic reconnection, producing a hard spectral slope $s \simeq 1$ up to $\gamma - 1 \simeq \sigma$; (ii) may experience very limited acceleration efficiency on scales $l< l_b \ll \lc$ where $\delta B \ll B$, but can still be energized with $s \simeq 1$ through mechanisms such as curvature acceleration; (iii) if particles interact for a sufficiently long time with structures where $\delta B \sim B$, they can achieve an overall slope of $s = 2$ in order to satisfy $\mathcal{L}(\gamma) \leq \mathcal{L}_b$; and finally, (iv) the distribution of confined particles is determined by Equation~\eqref{eq:slope} as $\rl$ approaches $\lc$. An additional escaping particle population is expected, described by Equation~\eqref{eq:escape}.
Note that if intermittent structures are prevalent at scales $l \ll \lc$, the characteristic scale $l_b$ shifts towards $l_b \to \sigma m c^2 / (e B \lc)$, maintaining a spectral slope of $\simeq 2$ from $\gamma \sim \sigma$ onward.

We define the scale at which the probability of remaining in the accelerating region decreases as the point where the spectral slope deviates by $\Delta s = 0.2$ from the canonical $s=2$ in Equation~\eqref{eq:slope}. This condition is satisfied at $l/\lc \sim \big[{\ln \mathcal{G} } \sqrt{{\sigma}/({1+\sigma})}/10\big]^{1/r}$. Consequently, intermediate scales with $s = 2$ can exist in realistic pair-plasma astrophysical environments if,
\begin{equation}\label{eq:lcB}
    \frac{\lc}{\rm cm} \frac{B}{\rm G} \gg \rm a \text{ } few \times \frac{10^6}{\sigma} 
\end{equation}
for $r\simeq 0.3$. For proton-electron plasmas or pair-plasma containing a non-negligible fraction of protons, this requirement translates to  $({\lc}/{\rm 10^9 cm}) ({B}/{1 \rm G}) \gg {7}/{\sigma}$. These requirements are met in jets, coronae, and magnetospheres.

The extent of the intermediate scales determines whether the system ultimately produces a power-law slope of 2. A key question, therefore, is whether Fermi processes are viable, or more specifically, whether intermittent structures are likely to emerge at these scales. Within our current PIC setup, the direct investigation of acceleration in this regime is constrained by the limited dynamical range. However, such intermittent structures could persist in magnetized regimes at low $\rl / \lc$ scales, distinct from scenarios where turbulence decays due to plasma effects \citep[e.g.,][]{lemoine+24}. This distinction is crucial, as turbulence decay in high-$\sigma$ environments remains uncertain, given that magnetic field energy is comparable to or exceeds plasma energy. Alternative acceleration mechanisms, such as magnetic reconnection, may become dominant while still leading to a similar spectral outcome \citep{mbarek+24}.

\section{Astrophysical Implications}

The arguments above provide an estimate of the particle distribution at the largest achievable Lorentz factors in plasma systems with magnetization $\sigma \geq 0.1$, once $\lc$, $\sigma$, $\sigma_e$, and $n_e$ are specified. This makes the framework applicable to jets, turbulent high-$\sigma$ magnetospheres, and black-hole coronae. As a concrete example, we apply it below to the Seyfert galaxy NGC~1068 and its potential neutrino emission \citep{IceCube-NGC1068,padovani+24}. This application is also motivated by several recent works connecting the neutrino signal from NGC~1068 to stochastic acceleration in turbulent black-hole coronae \citep{murase+20b,mbarek+24,fiorillo+24b,lemoine+25,lebihan+26}.

From the above considerations, we can express the energy-dependent proton spectrum in the corona as, $ \gamma \frac{dn}{d \gamma} \simeq  \bar{n}_p \left(\gamma_p/\sigma_p\right)^{-s+1}$, where $\gamma_p > \sigma_p$, $s$ are set by Equation~\eqref{eq:slope}, and $\bar{n}_p$ is the bulk density of the corona at $\gamma_p \simeq \sigma_p$.
In this context, low-density ``bursts'' of accelerated protons injected into the corona \citep{mbarek+24} via various mechanisms \citep[e.g.,][]{chashkina+21,mbarek+22,ripperda+22} could help pre-accelerate particles toward the maximum energy permitted by the available power at a scale $l < \lc$, at least for $\gamma < \gamma_{\rm max}$. However, they are not strictly necessary to produce NGC~1068's neutrinos if intermittent structures are present at scales $l \ll \lc$, providing the required acceleration.

\begin{figure}
	\centering
	\includegraphics[width=0.48\textwidth,clip=True,trim= 0 0 0 0]{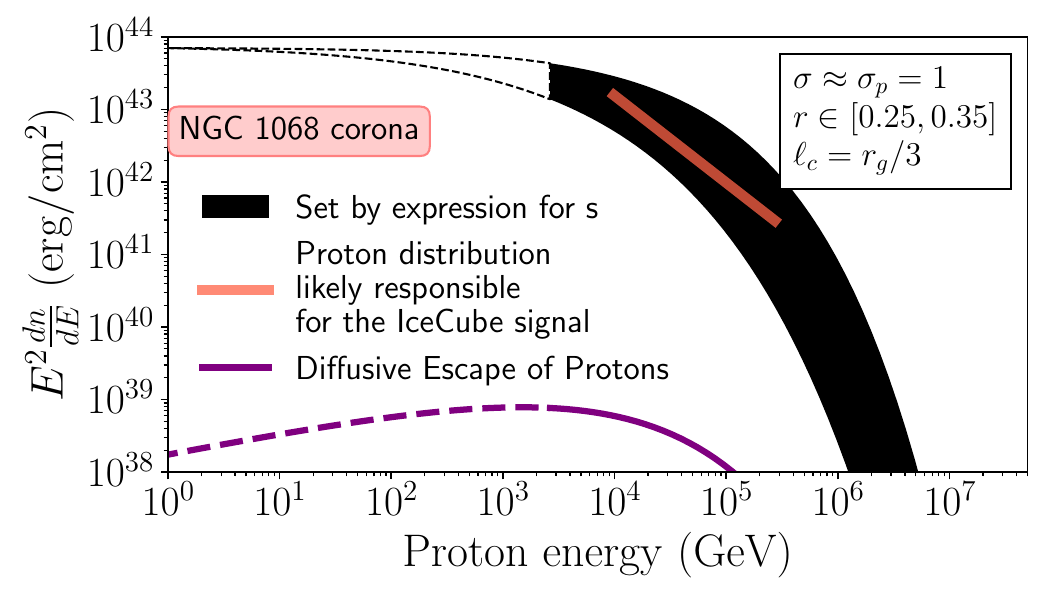}
	\caption{Spectrum of protons in the corona of NGC~1068 based on Equation~\eqref{eq:nptot} for observationally-motivated physical parameters. We find good agreement with the expected proton spectrum associated with NGC~1068's neutrino flux. The dashed lines are shown for normalization purposes and do not represent components of the obtained particle spectrum. The purple line denotes the escaping population of particles, i.e., diffusive coronal outflow.
    }
	\label{fig:astro}
\end{figure}

In the following, we express $n_p$ as a function of the coronal X-ray luminosity $\lx$, gravitational radius $r_g$, magnetization $\sigma$, and turbulent coherence length $\lc$, as an application of Equation~\eqref{eq:slope}. The bulk proton density $\bar{n}_p$ can be related to $\lx$ and $r_g$ by using the energetics of the optically thin corona. We first write the bulk electron density as $\bar{n}_e \simeq \tau/(\sigma_T r_c)$, where $\tau\simeq1$ \citep[e.g.,][]{rybicki+79,fabian+15,beloborodov17} is the optical depth and $\sigma_T$ is the Thomson cross-section. For the proton density, we use $\bar{n}_p\sigma_p=\sigma_e(m_e\bar{n}_e)/m_p=(2\ell/\tau)(U_{\rm B}/U_{\rm x})(m_e/m_p)\bar{n}_e$, where $\ell=\sigma_T U_{\rm x}r_c/m_ec^2$ is the radiative compactness, and $U_{\rm x}=\lx/4\pi c r_c^2$ and $U_B=B^2/8\pi$ are the X-ray and magnetic energy densities. This normalization corresponds to the limit in which an order-unity fraction of the coronal turbulent magnetic power is transferred to ions \citep{groselj+26}. In a turbulent corona scenario, $U_{\rm x}\sim2U_B$ \citep{groselj+24}, so we can rewrite $\bar{n}_p\simeq U_{\rm x}/(\sigma_p m_pc^2)$, and,
\begin{equation}\label{eq:nptot}
	 \gamma_p^2 \frac{dn_p}{d \gamma_p} \simeq \frac{10^{-3}\lx}{ \sigma_p m_p c^3 r_g^2} \left(\frac{\gamma_p}{\sigma_p}\right)^{-s+2}
\end{equation}
where the coronal size is consistently found to be $r_c \simeq 10 r_g$ \citep[e.g.,][]{dai+10,fabian12,fabian+15,wilkins+21}.

The slope of the distribution of accelerated protons, $s$, is described by Equation~\eqref{eq:slope} and requires an estimate of the skin depth in the corona, $d_e = \sqrt{{m_e c^2}/{4 \pi \bar{n}_e e^2}}$, $\sigma_{e}$, $\sigma_p$, and $\lc$. The skin depth $d_e$ only depends on $\bar{n}_e$, and thus can be calculated assuming the size of the corona. The magnetization $\sigma_{e}=B^2/(4\pi \bar{n}_e m_e c^2)=(2\ell/\tau) (U_B/U_{\rm x})$, where the compactness $\ell \approx 10$ for NGC~1068 \citep{mbarek+24}. We recover $\sigma_{e}\approx  \ell/\tau$, for the corona powered by magnetized turbulence.
As for $\sigma_p$, it is more challenging to extract directly from observations. However, if the proton spectral slope and the coherence length of the turbulent medium are known, $\sigma_p$ can in principle be inferred \citep[e.g.,][]{mbarek+24}.

In Figure~\ref{fig:astro}, we show the expected proton spectrum in NGC~1068 derived from Equation~\eqref{eq:nptot} for a reasonable range of $r$ values. We then compare this prediction with the proton spectrum required to account for the $\gtrsim 10$ TeV neutrinos associated with NGC~1068 in the relevant energy range \citep[e.g.,][]{mbarek+24}. Treating $\lc$ as a free parameter, we find good agreement for $\lc = r_g/3$, consistent with expectations from PIC simulations of turbulence \citep{sironi+23}, which suggest that the coherence length is roughly $1/20$ of the system size.
These results highlight the applicability of our framework to magnetized environments such as black hole coronae. 

\balance
\section{Conclusions}
We present an analytical expression for the slope of the nonthermal particle distribution in magnetically-dominated turbulent plasma (see Equation~\eqref{eq:slope}). This formalism only depends on macro-scale properties of the turbulent environment, including the magnetization $\sigma$, coherence length $\lc$, and skin depth $d_e$.
Our results agree with spectral features of PIC driven turbulence simulations with magnetization $\sigma \geq 0.1$. Testing this formalism against the observationally motivated plasma properties of NGC~1068, we find that the predicted proton distribution aligns well with the proton population inferred from the source's neutrino emission at $\gtrsim $TeV energies. Our main conclusions include:

\begin{itemize}
    \item The slope ($s \to 2$) is an attractor for $\sigma \ge 0.1$ when regions with $\delta B/B \sim 1$ have a non-negligible filling factor. Broken power laws are not strictly necessary, as the spectrum naturally steepens as $\rl \to \lc$ (Equation~\eqref{eq:slope}).
    \item The exact form of the expression for the acceleration time in Eq.~\eqref{eq:tacc} depends on the energization channel and the pitch angle distribution. Here we focus on mirror acceleration, which is expected to dominate when $\delta B/B \sim 1$. A comprehensive treatment that quantifies the dependence on $\delta B/B$, especially for $l \ll \lc$, is left for future work.
    \item The steepening of the turbulent spectrum reflects the reduced efficiency of particle interactions with turbulent fluctuations at scales comparable to the Larmor radius, $\rl$. Scales for which $s\to2$ have particle power comparable to the available magnetic energy $\sim B^2/8\pi$.
    \item The diffusively escaping population is generically harder than the confined one by $\Delta s \simeq r$ as in Eq.~\eqref{eq:escape}. 
    \item Intermittent magnetic structures likely drive variability. Temporal changes in the index of the tail of the  curvature distribution, $\mathcal{P}_\kappa$, change the exponent $r$ and therefore $s$.
    \item The formalism developed here applies broadly to magnetized astrophysical environments, including jets, turbulent large-$\sigma$ magnetospheres, and coronae, and provides a means to infer turbulent plasma properties.
    \item Applied to the corona of NGC~1068, our formalism suggests that turbulent acceleration can account for a population of high-energy protons. The precise spectral shape at GeV–TeV energies remains uncertain, as it likely depends on the degree of intermittency and additional acceleration processes beyond those considered here.
    \item Future work should extend the generalized tests presented here by following individual particle trajectories to directly quantify how the interaction probability scales at large Larmor radii, and to assess whether this scale-dependent suppression can explain the observed spectral steepening.
\end{itemize}

\begin{acknowledgments}
R.M. is supported by a Lyman Spitzer Fellowship at Princeton University. D.G.~is supported by the Research Foundation--Flanders (FWO) 
Senior Postdoctoral Fellowship 12B1424N. We would like to thank Martin Lemoine for kindly providing comments on the manuscript. We would like to thank Lorenzo Sironi and Philipp Kempski for useful conversations.
\end{acknowledgments}
{\software{
\textsc{Tristan-MP v2} \citep{tristanv2},
matplotlib \citep{Hunter:2007},
numpy \citep{harris2020array},
scipy \citep{2020SciPy-NMeth}}

\balance

\bibliography{Total}
\bibliographystyle{aasjournalv7}

\appendix

\section{Dependence of $\delta B/B$ on curvature}\label{app:deltaB}

In this section, we plot the magnetic-field strength $B(\kappa)/B_{\mathrm{rms}}$ as a function of the field-line curvature for our fiducial simulation with $\sigma_p = 2$. The quantity $B(\kappa)$ represents the mean magnetic-field strength associated with a given field-line curvature $\kappa$ and is obtained by coarse-graining the magnetic field at a specified scale $l$ and binning the local field magnitudes according to their corresponding curvature values. In contrast, $B_{\mathrm{rms}}$ denotes the root-mean-square magnetic-field strength, defined as $B_{\mathrm{rms}} = \langle |\mathbf{B}|^2 \rangle^{1/2}$, which provides a global measure of the average magnetic-field amplitude within the domain. The resulting profile $\langle B(\kappa)\rangle / B_{\mathrm{rms}}$ quantifies how the magnetic-field intensity varies with curvature, revealing the correlation between magnetic-field strength and the degree of field-line bending. Figure~\ref{fig:deltaB} shows that this relationship is approximately constant over an extended range of scales, consistent with results from 3D large-amplitude turbulence with $\delta B/B \approx 1$ \citep{golant+25}.

\begin{figure}[htb]
	\centering
\includegraphics[width=0.48\textwidth,clip=false,trim= 0 0 0 0]{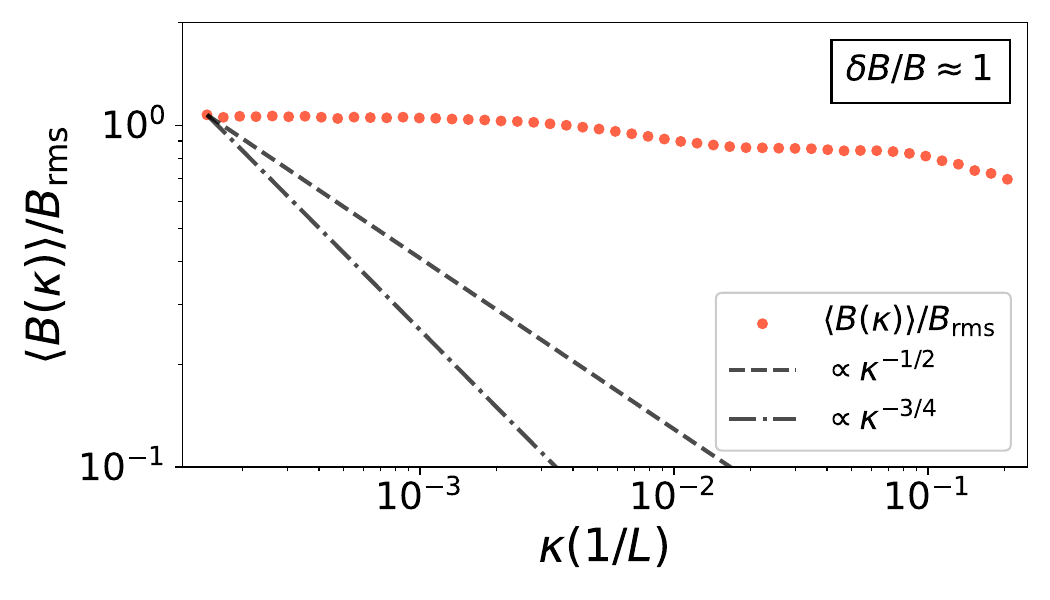}
	\caption{Relationship between the normalized magnetic-field strength and field-line curvature. The curve shows the mean magnetic-field amplitude $\langle B(\kappa)\rangle / B_{\mathrm{rms}}$ is roughly constant as a function of curvature $\kappa$. Reference power-law trends ($B \propto \kappa^{-1/2}$ and $B \propto \kappa^{-3/4}$) are overplotted for comparison.}
	\label{fig:deltaB}
\end{figure}

\section{Probability distribution of curvature for $\sigma_p = 5$}\label{app:intermit}

\begin{figure}[htb]
	\centering
\includegraphics[width=0.48\textwidth,clip=false,trim= 0 0 0 0]{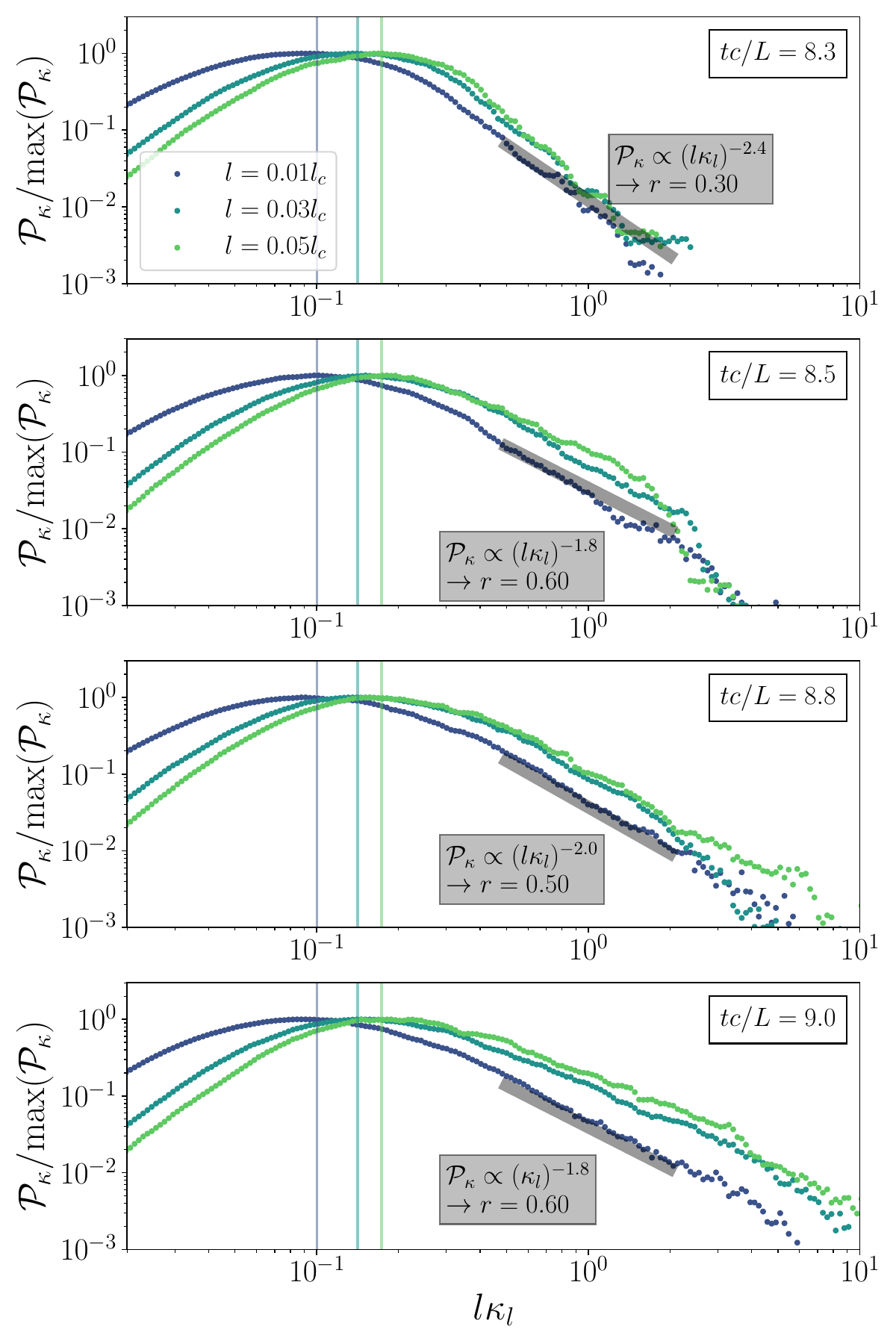}
	\caption{Statistics of the curvature scales $\langle l \kappa_l \rangle$ in a 2D box of size $L/d_e = 3200$ and $\sigma_p = 5$ at different times.
    The curvature $\langle \kappa_l \rangle$ is defined in Equation~\eqref{eq:kappa}. The distributions show tails that scale as $\mathcal{P}_\kappa \propto (l \kappa)^{-\alpha}$ for $\langle l \kappa_l \rangle \gtrsim 1$. Importantly, a harder slope is sustained for a longer time resulting in an average $r \approx 0.5$.}
	\label{fig:s5-Intermit}
\end{figure}

We show in Figure~\ref{fig:s5-Intermit} the statistics of the curvature scales for a simulation initialized with $\sigma_p = 5$. We find that harder $\mathcal{P}_\kappa$ slope are sustained for a longer time resulting in an average $r \approx 0.5$. The impact of the box dimensions and $\delta B/ B$ on intermittent structures setting the value of $r$ should be investigated.

\end{document}